\documentclass[12pt]{article}
\usepackage{amsmath,amssymb,mathrsfs,amsbsy,latexsym,amsfonts,amsthm, ascmac}
\usepackage{fancyhdr}
\usepackage{braket}
\usepackage{authblk}
\usepackage{graphicx}
\usepackage{color}
\usepackage{physics}
\usepackage{hyperref}
\hypersetup{
    unicode=false,          % non-Latin characters in Acrobat’s bookmarks
    pdftoolbar=true,        % show Acrobat’s toolbar?
    pdfmenubar=true,        % show Acrobat’s menu?
    linktocpage=true,       % hyperlink by page number in contents
    pdffitwindow=false,     % window fit to page when opened
    pdfstartview={FitH},    % fits the width of the page to the window
    pdftitle={two-point},    % title
    pdfnewwindow=true,      % links in new PDF window
    colorlinks=true,       % false: boxed links; true: colored links
    linkcolor=blue,          % color of internal links (change box color with linkbordercolor)
    citecolor=blue,        % color of links to bibliography
    filecolor=blue,      % color of file links
    urlcolor=blue           % color of external links
}
\usepackage[top=25truemm,bottom=28truemm,left=22truemm,right=22truemm]{geometry}
\numberwithin{equation}{section}

\newcommand{\nn}{\nonumber\\}
\newcommand{\vp}{\vec{p}}

\newcommand{\hp}{\hat{p}}

\newcommand{\cO}{\mathcal{O}}
\newcommand{\cA}{\mathcal{A}}
\newcommand{\tout}{\text{out}}
\newcommand{\tin}{\text{in}}
\usepackage{epsf}

\title{\bf {\fontsize{19pt}{19pt}\selectfont Celestial two-point functions 
and rectified dictionary}}
\author[1]{
	Hideo~Furugori\thanks{\tt h-furugori(at)gauge.scphys.kyoto-u.ac.jp}
}
\author[2]{
	Naoki~Ogawa\thanks{
	\tt naoki.ogawa(at)yukawa.kyoto-u.ac.jp}
}
\author[1]{
	Sotaro~Sugishita\thanks{
	\tt sotaro(at)gauge.scphys.kyoto-u.ac.jp}
	\vspace{5mm}
}
\author[2]{
	Takahiro~Waki\thanks{
	\tt takahiro.waki(at)yukawa.kyoto-u.ac.jp}
}
\affil[1]{\it\normalsize Department of Physics, Kyoto University, Kyoto 606-8502, Japan}
\affil[2]{\it\normalsize Center for Gravitational Physics and Quantum Information,\protect\\
Yukawa Institute for Theoretical Physics, Kyoto University, \protect\\
Kitashirakawa Oiwakecho, Sakyo-ku, Kyoto 606-8502, Japan}
\date{ }

\begin{document}
\maketitle
\thispagestyle{fancy}
\renewcommand{\headrulewidth}{0pt}

\begin{abstract}
A naive celestial dictionary causes massless two-point functions to take the delta-function forms in the celestial conformal field theory (CCFT). 
We rectify the dictionary, involving the shadow transformation so that the two-point functions follow the standard power-law.
In this new definition, we can smoothly take the massless limit of the massive dictionary.
We also compute a three-point function using the new dictionary and discuss the OPE in CCFT.
\end{abstract}

\newpage
\thispagestyle{empty}
\setcounter{tocdepth}{2}

\setlength{\abovedisplayskip}{12pt}
\setlength{\belowdisplayskip}{12pt}

\tableofcontents
\newpage
\section{Introduction}
The holographic principle \cite{tHooft:1993dmi, Susskind:1994vu} 
states the duality between a quantum gravity in some spacetime and a non-gravitational theory in a lower-dimensional spacetime.
Representative examples are the AdS/CFT correspondence \cite{Maldacena:1997re,Gubser:1998bc, Witten:1998qj}, and also the dS/CFT correspondence \cite{Strominger:2001pn, Maldacena:2002vr}.
Recently, many researchers have sought to establish a holographic correspondence in an asymptotic flat spacetime, celestial holography \cite{Pasterski:2016qvg, Cardona:2017keg, Strominger:2017zoo, Pasterski:2017kqt, Pasterski:2017ylz, Donnay:2020guq, Ogawa:2022fhy}. 
Celestial holography is a conjecture about a correspondence between a quantum gravity in a four-dimensional asymptotically flat spacetime and a conformal field theory on a two-dimensional sphere (celestial sphere $\mathcal{CS}^2$). 
This holography relies on the fact that the symmetry of a gravitational theory on asymptotically flat spacetime seems to contain the Virasoro symmetry \cite{Kapec:2014opa, Kapec:2016jld}.

The central object of the celestial holography is the relation between scattering amplitudes of a quantum theory in a four-dimensional asymptotically flat spacetime and correlation functions of a conformal field theory on the celestial sphere (celestial conformal field theory; CCFT), which is schematically expressed as follows:
\begin{align}
    \bra{\text{out}}\mathcal{S}\ket{\text{in}}=\langle \mathcal{O}_{\text{out}}\cdots\mathcal{O}_{\text{in}} \rangle_{\mathcal{CS}^2}.
\end{align}
These operators $\mathcal{O}$ in CCFT are constructed from field operators $\phi(X)$ in four-dimensional spacetime so that $\mathcal{O}$ transform appropriately under the two-dimensional conformal transformations.
This construction has been achieved by extracting components for a specific basis called the conformal primary wave function $\Phi_{\Delta}\left(X; z\right)$ (see e.g. \cite{Donnay:2020guq, Pasterski:2021rjz, Raclariu:2021zjz}) as
\begin{align}
    \mathcal{O}_{\Delta}(z) := \left(\Phi_{\Delta}\left(X; z\right),\phi\left(X\right)\right)_{\text{KG}}.
\end{align} 
However, the proposed dictionary in the literature leads to a peculiar form for the two-point functions for massless fields. 
Indeed, the two-point functions take the following delta-functional behavior if we use the proposed dictionary:\footnote{This fact is already noted in \cite{Crawley:2021ivb}. It is also remarked in \cite{Chang:2022jut} that massless four-point functions also take the delta function forms.}
\begin{align}
\left\langle \mathcal{O}_{\Delta}(z_1)\mathcal{O}_{\Delta}(z_2)\right\rangle_\text{naive} \propto \delta^{(2)}(z_1-z_2).
\end{align}
On the other hand, three-point functions take the standard forms determined by the conformal transformation (e.g. \cite{Raclariu:2021zjz, Chang:2022seh} ). 
It looks hard to reconcile these behaviors with the operator product expansion (OPE). 

In this paper, we rectify the dictionary by adopting alternative conformal wave functions for massless incoming particles. 
This rectification is equivalent to interpreting operators constructed from the conventional massless conformal wave functions as the conformal shadow operators for incoming particles.
In our dictionary, the celestial two-point functions follow the standard power-law:
\begin{align}
    \left\langle \cO_{\Delta} (z_1)\cO_{\Delta}(z_2)\right\rangle_\text{new} \propto 
    \frac{1}{|z_1-z_2|^{2\Delta}}.
\end{align}
Furthermore, our rectification does not affect the form of three-point functions. 
% \begin{align}
%     \left\langle \cO^-_{1+i\lambda_1(new)} (z_1)\cO^+_{1+i\lambda_2(new)}(z_2)\right\rangle=
%     \frac{ \delta(\lambda_1-\lambda_2)}{|z_1-z_2|^{2(1+i\lambda_1)}}.
% \end{align}

Here we note that a similar idea is proposed in \cite{Chang:2022jut}. However, their approach is different from ours. In \cite{Chang:2022jut} the shadow transformation is taken for all the particles including outgoing particles, and then two-point functions still result in the delta functions. 
It is important to adopt the shadow transformation only for incoming particles to obtain the power-law for the two-point functions.\footnote{This asymmetry between outgoing and incoming may be related to the antipodal relation discussed in the context of the asymptotic symmetry (see, e.g., \cite{He:2014cra}), although we leave further discussions for future works.} 

Our rectification is supported by the fact that our altered choice of the conformal wave functions for massless fields is smoothly connected to the massive wave functions (used in the literature) in the massless limit.
It is further supported by the crossing symmetry in the context of scattering amplitudes.

This paper is organized as follows.
In Sec.~\ref{Sec2}, we propose a rectified celestial dictionary. 
By considering the two-point scattering amplitudes for one particle passing through, we redefine operators in CCFT to achieve a desirable form for the two-point functions. Specifically, we alter the treatment of incoming particles. 
We consider the massless limit and see the consistency of the dictionary for massless and massive fields. We further discuss this modification from the viewpoint of the crossing symmetry.
In Sec.~\ref{Sec3}, we discuss the validity of our definitions using OPE. In addition, we reevaluate the celestial three-point correlator for one massive and two massless particles, which was computed previously, using our new definitions, confirming that the form of the three-point function is appropriate. Sec.~\ref{Sec4} is devoted to a summary.

\paragraph{Notation}
Here we summarize the notation used in this paper. 

Massless on-shell momenta $p^\mu$ will be parameterized by $\omega\, (0\leq \omega <\infty)$ and angle coordinates, which we will use complex coordinates $z \in \mathbb{C}$, as, 
\begin{align}
\label{m0-p}
    p^\mu=\omega q^{\mu}(z), \qquad
    q^{\mu}(z)=(1+z\bar{z},z+\bar{z},i(\bar{z}-z),1-z\bar{z}).
\end{align} 
We define the integral measure on $\mathbb{C}$ as $ \int d^2 z=2\int d\, \mathrm{Re}[z]\, d\, \mathrm{Im}[z]$ and the delta function $\delta^{(2)}(z_1-z_2)\equiv \frac{1}{2}\delta(\mathrm{Re} z_1-\mathrm{Re} z_2)\delta(\mathrm{Im}z_1-\mathrm{Im}z_2)$ as in \cite{Polchinski:1998rq}.

For massive on-shell momenta $p^\mu$ with mass $m$, we introduce $\hp^\mu:=m^{-1}p^\mu$. 
$\hp^\mu$ satisfies $\hp^2=-1$ and it represents the embedding coordinates of the three-dimensional unit hyperbolic space $H^{3}$ to $\mathbf{R}^{1,3}$. 
$\hp^\mu$ can be parameterized by the Poincare coordinates $(y,z)$ as
\begin{align}
   \hp^\mu(y,z)=\frac{1}{2y}(1+y^2+z\bar{z}, z+\bar{z},i(\bar{z}-z),1-y^2-z\bar{z}).
\end{align}
We define the integral measure on $H^{3}$ as
\begin{align}
    \int [d\hp]:=\int\frac{d^3 \hp^i}{\hp^0}=\int\frac{dy\,d\, \mathrm{Re}[z]\, d\, \mathrm{Im}[z] }{y^3}=\int\frac{dy\,d^2z }{2y^3}.
\end{align}

%%%%%%%%%%%%%%%
%%%%%%%%%%%%%%%
\section{Rectified holographic dictionary and two-point functions}\label{Sec2}
In the celestial holography, the following relation between scattering amplitudes of massless fields and correlation functions is proposed (e.g. \cite{Pasterski:2021rjz}):
\begin{align}
    \ev{\prod_{i=1}^{n}\mathcal{O}_{\Delta_i}}=\qty[\prod_{j=1}^{n}\int d\omega_j~\omega_j^{\Delta_i-1}]\mathcal{A}(p_1,\cdots,p_{n}).
\end{align}
It has a problem, especially in the two-point functions as follows. 
The massless two-point scattering amplitude is given by
\begin{align}
    \mathcal{A}(p_1,p_2)=(2\pi)^3 2 p_1^0 \delta^{(3)}(\vec{p}_1-\vec{p}_2).
\end{align}
From it, we can calculate the celestial two-point correlators as 
\begin{align}
    \ev{\mathcal{O}_{\Delta_1=1+i\lambda_1}(z_1)\mathcal{O}_{\Delta_2=1+i\lambda_2}(z_2)}&=\prod_{i=1,2}\qty(\int d\omega_i~\omega_i^{i\lambda_i})\mathcal{A}(p_1,p_2)\nonumber\\
    &=(2\pi)^4\delta(\lambda_1+\lambda_2)\delta^{(2)}(z_1-z_2).
\end{align}
Thus, the two-point functions for massless scalar fields show a delta-function behavior. The conformal symmetry may allow this delta-function behavior of the two-point functions. 
However, as we will discuss in Sec.~\ref{Sec3}, it is problematic for the property of the CCFT. 
Therefore, in this section, we rectify the dictionaries for celestial holography in order to make two-point functions an appropriate form. Our rectification is natural in terms of the massless limit of massive dictionaries and a crossing symmetry for scattering amplitudes. 

\subsection{Massless scalar}\label{subsec:m0}
The celestial holographic dictionary gives us a translation between scattering amplitudes in the Minkowski space and the correlation functions on the CCFT. 
In this subsection, we consider massless scalar fields. 
For the scattering amplitude  
$\cA(\{p^\text{out}_i\},\{p^\text{in}_j\})$, we propose the following dictionary:\footnote{The phase factors are fixed so that they agree with the massless limit of the massive dictionary as we will see.} 
\begin{align}
   &\left\langle \prod_i^m \cO^-_{\Delta_i} \prod_j^n \widetilde{\cO}^+_{\Delta_j}\right\rangle := \prod_{i=1}^m \prod_{j=1}^n\int d\omega_i \frac{-i\omega_i^{i\lambda_i}}{\sqrt{(2\pi)^3\lambda_i}}\int d\omega_j \frac{i\omega_j^{-i\lambda_j}}{\sqrt{(2\pi)^3\lambda_j}}i\cA(p^\text{out}_1,...,p^\text{out}_m; p^\text{in}_1,...,p^\text{in}_n).
   \label{massless-dict-amp}
\end{align}
Here conformal dimension $\Delta_i$ takes principal series $\Delta_i\in 1+i \lambda_i$ $(\lambda_i \in \mathbb{R})$. 
It is essentially the Mellin transformation of the scattering amplitudes through the parameterization \eqref{m0-p} as proposed in the literature. 
The difference from the literature is that we interpret the operators corresponding to the incoming particles as the shadow-transformed operators $\widetilde{\cO}_\Delta$ defined by (for details of shadow operators, see, e.g., \cite{Simmons-Duffin:2012juh}) 
\begin{align}
    \widetilde{\cO}_\Delta(z):=\frac{\Delta-1}{2\pi}\int d^2 z'\frac{1}{|z-z'|^{2(2-\Delta)}}\cO_{\Delta}(z').
    \label{shadow}
\end{align}
The reason why we take this rectification of the dictionary is that the two-point functions of (non-shadow) operators take the conventional form in CFT as we see below. 

Let us consider the two-point functions obtained from the above dictionary. 
The scattering amplitude is  trivial as
\begin{align}
    \cA(p_1; p_2)=(2\pi)^3 2 p_1^0 \delta^{(3)}(\vec{p}_1-\vec{p}_2).
\end{align}
Using the parameterization \eqref{m0-p}, the delta function can be rewritten as:
\begin{align}
    \delta^{(3)}(\vec{p}_1-\vec{p}_2)=\frac{1}{2\omega_1^2(1+|z_1|^2)}\delta(\omega_1-\omega_2)\delta^{(2)}(z_1-z_2).
\end{align}
Thus, we have
\begin{align}
    \cA(p_1; p_2)= \frac{(2\pi)^3}{\omega_1}\delta(\omega_1-\omega_2)\delta^{(2)}(z_1-z_2).
\end{align}
Using the dictionary \eqref{massless-dict-amp}, the obtained two-point function is 
\begin{align}
\label{oto}
     \left\langle \cO^-_{1+i\lambda_1} (z_1)\widetilde{\cO}^+_{1+i\lambda_2}(z_2)\right\rangle=\frac{2\pi i}{\lambda_1}\delta(\lambda_1-\lambda_2)\delta^{(2)}(z_1-z_2).
\end{align}
Performing the inverse shadow transformation on $\widetilde{\cO}^+(z_2)$, we can obtain the two-point function of non-shadow operators as\footnote{
We use the formula
\begin{align}
    \int d^2 z \frac{1}{|z-z'|^{2(1+i\lambda)}|z-z''|^{2(1-i\lambda)}}=\frac{4\pi^2}{\lambda^2}\delta^{(2)}(z'-z''),
\end{align}
and then we have
\begin{align}
    \widetilde{\widetilde{\cO}}_{1+i\lambda}(z)=\frac{-i\lambda}{2\pi}\frac{i\lambda}{2\pi}\int d^2 z' d^2 z'' \frac{1}{|z-z'|^{2(1+i\lambda)}|z'-z''|^{2(1-i\lambda)}}\cO_{1+i\lambda}(z'')=\cO_{1+i\lambda}(z).
\end{align}
}
\begin{align}
\label{m0-2pt-power}
     \left\langle \cO^-_{1+i\lambda_1} (z_1)\cO^+_{1+i\lambda_2}(z_2)\right\rangle=
    \frac{ \delta(\lambda_1-\lambda_2)}{|z_1-z_2|^{2(1+i\lambda_1)}}.
\end{align}
Therefore, the above dictionary leads to the standard power-law form of the conformal two-point functions.
If we do not interpret the operators corresponding to the incoming particles as the shadow operators, the two-point functions become the delta function $\delta^{(2)}(z_1-z_2)$ as \eqref{oto}.

The dictionary can be interpreted as the relation between the creation-annihilation operators in the Minkowski space and the operators in the correlation function as follows:
\begin{align}
     \cO_{1+i\lambda}^-(z)=\int^\infty_0 d\omega \frac{-i\omega^{i\lambda}}{\sqrt{(2\pi)^3\lambda}}a_\tout(\omega q(z)), \quad \widetilde{\cO}_{1+i\lambda}^+(z)=\int^\infty_0 d\omega \frac{i\omega^{-i\lambda}}{\sqrt{(2\pi)^3\lambda}}a_{\tin}^{\dagger}(\omega q(z)).
     \label{O-a}
\end{align}
Here, the bulk field $\phi(x)$ is related to the asymptotic fields $\phi^{\text{in/out}}$ in the Heisenberg picture:
\begin{align}
    \phi(x) &\to \sqrt{Z}\phi^\tout(x) \qquad (t\to +\infty), \\
    \phi(x) &\to \sqrt{Z}\phi^\tin(x) \qquad (t\to -\infty),
\end{align}
where they have the mode expansion
\begin{align}
    \phi^{\tout/\tin}(x)=\int \frac{d^3 p}{(2\pi)^3(2p^0)}\left[a_{\tout/\tin}(p)e^{ip\cdot x}+a_{\tout/\tin}^{\dagger}(p)e^{-ip\cdot x}\right].
\end{align}

For free fields, $a_\tin=a_\tout$, and we have
\begin{align}
\label{dag=til}
    (\cO_{1+i\lambda}^-(z))^\dagger&=\int^\infty_0 d\omega \frac{i\omega^{-i\lambda}}{\sqrt{(2\pi)^3\lambda}}a^\dagger(\omega q(z))=\widetilde{\cO}_{1+i\lambda}^+(z).
\end{align}
Thus, the Hermitian conjugation with respect to the Klein-Gordon inner product in the Minkowski space corresponds to the shadow transformation on the celestial sphere.
This is different from \cite{Crawley:2021ivb} where the Hermitian conjugation with respect to ``the shadow product''  (which is different from the usual Klein-Gordon product) corresponds to the two-dimensional shadow transformation. 
Accordingly, in \cite{Crawley:2021ivb}, it is argued that $\widetilde{\cO}$ is the conjugation of $\cO$ with respect to the BPZ inner product because $\langle \cO \widetilde{\cO} \rangle$ follows the power-law form in their formulation. 
In our formulation, $\widetilde{\cO}$ is not the conjugation of $\cO$ w.r.t. the BPZ inner product because $\langle \cO \widetilde{\cO} \rangle$ is a delta function.

The relation \eqref{O-a} can be rewritten as 
\begin{align}
     \cO_{1+i\lambda}^-(z)=(\varphi^-_{1+i\lambda}(x;z),\phi^\tout(x))_{\text{KG}}, \quad 
    \widetilde{\cO}_{1+i\lambda}^+(z)=(\widetilde{\varphi}^+_{1+i\lambda}(x;z),\phi^\tin(x))_{\text{KG}},
    \label{O-m0-KG}
\end{align}
where  the symbol $(,)_{\text{KG}}$ denotes the Klein-Gordon inner product defined by
\begin{align}
(\phi_1,\phi_2)_{\text{KG}}:=-i\int d^3\Sigma^{\mu}(\phi_1 \partial_\mu \phi^\ast_2-(\partial_\mu \phi_1) \phi_2^\ast).
\end{align}
$\varphi^{-}_{1+i\lambda}(x,z)$ and $\widetilde{\varphi}^{+}_{1+i\lambda}(x,z)$ are massless conformal primary wave functions defined by
\begin{align}
\label{vphi-int}
   \varphi^-_{1+i\lambda}(x;z):= \int^\infty_0 d\omega \frac{i\omega^{i\lambda}}{\sqrt{(2\pi)^3\lambda}}e^{-i\omega q(z)\cdot x}\,,\quad\,
   \widetilde{\varphi}^+_{1+i\lambda}(x;z):= \int^\infty_0 d\omega \frac{i\omega^{-i\lambda}}{\sqrt{(2\pi)^3\lambda}}e^{i\omega q(z)\cdot x}.
\end{align}
Computing the integrals, we obtain
\begin{align} 
\label{vphi-}
 \varphi^-_{1+i\lambda}(x;z)&= -\frac{i\,e^{-\frac{\pi\lambda}{2}}\sqrt{\lambda}\Gamma(i\lambda)}{(2\pi)^\frac{3}{2}(-q(z)\cdot x)^{1+i\lambda}},
 \\
 \widetilde{\varphi}^+_{1+i\lambda}(x;z)&= -\frac{i\,e^{-\frac{\pi\lambda}{2}}\sqrt{\lambda}\Gamma(-i\lambda)}{(2\pi)^\frac{3}{2}(-q(z)\cdot x)^{1-i\lambda}}.
 \label{tvphi+}
\end{align}
Here, we note that this $\widetilde{\varphi}^+$ is defined so that it gives \eqref{O-m0-KG}, and is proportional to the standard conformal wave functions (non-shadow one) in, e.g., \cite{Pasterski:2021rjz, Raclariu:2021zjz}. 

From \eqref{O-m0-KG}, we also obtain the formula directly computing $\cO_{1+i\lambda}^+$ (non-shadow one) as 
\begin{align}
    \cO_{1+i\lambda}^+(z)=(\varphi^+_{1+i\lambda}(x;z),\phi^\tin(x))_{\text{KG}},
\end{align}
where $\varphi^+_{1+i\lambda}$ is an inverse shadow transformed function from $\widetilde{\varphi}^+_{1+i\lambda}$ defined by
\begin{align}
    \varphi^+_{1+i\lambda}(x;z)&:=\frac{-i\lambda}{2\pi}\int d^2 z'\frac{1}{|z-z'|^{2(1+i\lambda)}}\widetilde{\varphi}_{1+i\lambda}^+(x;z')
    \nn
    &= -\frac{i\,e^{\frac{\pi\lambda}{2}}\sqrt{\lambda}\Gamma(-i\lambda)(x^2)^{i\lambda}}{(2\pi)^\frac{3}{2}(-q(z)\cdot x)^{1+i\lambda}}.
    \label{vphi+}
\end{align} 
This $\varphi^+_{1+i\lambda}$ is a so-called shadow conformal wave function, up to a factor, in the literature (see \cite{Pasterski:2017kqt}). 
We have to use it for incoming particles to obtain the power-law two-point functions \eqref{m0-2pt-power}.

%%%%%%%%
\subsection{Massive scalar}
Next, we consider the massive scalars. 
Our dictionary for the massive scalars is the same as that in literature, up to multiplicative factors.
However, we will see that the massless limit of the conformal wave functions in this conventional dictionary reproduces our massless conformal wave functions for the rectified dictionary given in the previous subsection \ref{subsec:m0}.
This fact supports our rectified dictionary. 

The holographic dictionary for massive scalars is as follows: \begin{align}
   &\left\langle \prod_i^m \cO^-_{1+i\lambda_i} (z_i)\prod_j^n \cO^+_{1+i\lambda_j}(z_j)\right\rangle \nn
   &:= \prod_{i=1}^m \prod_{j=1}^n\int [d\hp_i]c(m_i, \lambda_i)G_{1+i\lambda_i}(\hp_i^\tout; z_i)\int [d\hp_j] c(m_j, \lambda_j)^{\ast}G_{1+i\lambda_j}(\hp_j^\tin; z_j)\,i\cA(p^\text{out}_1,...,p^\text{out}_m; p^\text{in}_1,...,p^\text{in}_n),
   \label{massive-dict-amp}
\end{align}
where $c(m, \lambda):=2^{-i\lambda}m^{1+i\lambda} \sqrt{\lambda}/(2\pi)^\frac{5}{2}$ which is chosen so that the two-point functions are normalized as we will see below.
Here, $G_\Delta(\hp;z)$ is a bulk-boundary propagator used in the Euclidean AdS$_3$/CFT$_2$ correspondence:
\begin{align}
    G_\Delta(\hat{p}(y,w);z)=\frac{1}{(-\hp(y,w)\cdot q(z))^{\Delta}}=\left( \frac{y}{y^2+|z-w|^2} \right)^{\Delta}.
\end{align}
We should note that the principal series of massive scalar fields is $\Delta\in 1+i\lambda$ where $\lambda>0$.
Up to the multiplicative factors, this dictionary is the same as that in literature (see, e.g., \cite{Pasterski:2017kqt}). 

To compare the massive dictionary with the massless one, we also write down the dictionary involving the shadow operator. The above dictionary \eqref{massive-dict-amp} leads to\footnote{
Note that 
\begin{align}
    \frac{-i\lambda}{2\pi}\int d^2 z' \frac{1}{|z-z'|^{2(1+i\lambda)}}G_{1-i\lambda}(\hp; z')=G_{1+i\lambda}(\hp; z).
\end{align}
} 
\begin{align}
   &\left\langle \prod_i^m \cO^-_{1+i\lambda_i} \prod_j^n \widetilde{\cO}^+_{1+i\lambda_j}\right\rangle \nonumber\\
   &= \prod_{i=1}^m \prod_{j=1}^n\int [d\hp_i]c(m_j, \lambda_j)G_{1+i\lambda_i}(\hp_i^\tout; z_i)\int [d\hp_j] c(m_j, \lambda_j)^{\ast} G_{1-i\lambda_j}(\hp_j^\tin; z_j)i\cA(p^\text{out}_1,...,p^\text{out}_m; p^\text{in}_1,...,p^\text{in}_n).
   \label{massive-dict-shadow}
\end{align}

Let us confirm that this definition leads to the standard form of conformal two-point functions. 
For the Minkowski two-point amplitude
\begin{align}
    \cA(p_1; p_2)=(2\pi)^3 2 p_1^0 \delta^{(3)}(\vp_1-\vp_2)=\frac{(2\pi)^3}{m^2} 2 \hp_1^0 \delta^{(3)}(\hp_1-\hp_2),
\end{align}
the above dictionary leads to
\begin{align}
 &\left\langle \cO^-_{1+i\lambda_1} (z_1)\cO^+_{1+i\lambda_2}(z_2)\right\rangle=
    \frac{ \delta(\lambda_1-\lambda_2)}{|z_1-z_2|^{2(1+i\lambda_1)}},
    \\
     &\left\langle \cO^-_{1+i\lambda_1} (z_1)\widetilde{\cO}^+_{1+i\lambda_2}(z_2)\right\rangle=\frac{2\pi i}{\lambda_1}\delta(\lambda_1-\lambda_2)\delta^{(2)}(z_1-z_2).
\end{align}
Thus, for massive scalars, the dictionary proposed in the literature leads to the standard form of the two-point functions.

As in the massless fields, we can relate the CCFT operator to the creation/annihilation operators as follows:
\begin{align}
     \cO_{1+i\lambda}^-(z)&= \int [d\hp]\frac{ 2^{-i\lambda} m^{1+i\lambda} \sqrt{\lambda}G_{1+i\lambda}(\hp; z)}{(2\pi)^\frac{5}{2}}a_\tout(m \hp),
     \\
    \widetilde{\cO}_{1+i\lambda}^+(z)&=\int [d\hp]\frac{2^{i\lambda} m^{1-i\lambda} \sqrt{\lambda}G_{1-i\lambda}(\hp; z)}{(2\pi)^\frac{5}{2}}a_{\tin}^{\dagger}(m \hp).
\end{align}
For free fields, $a_\tin=a_\tout$, we have
\begin{align}
    (\cO_{1+i\lambda}^-(z))^\dagger&=\int [d\hp]\frac{ m \sqrt{\lambda}G_{1-i\lambda}(\hp; z)}{(2\pi)^\frac{5}{2}}a^\dagger(m \hp)=\widetilde{\cO}_{1+i\lambda}^+(z).
\end{align}
This relation is the same as the massless case. 

The operator relation can be rewritten as\footnote{Similarly, we can define $\widetilde{\cO}_{1+i\lambda}^-(z)=(\widetilde{\Phi}^-_{1+i\lambda}(x;z),\phi^\tout(x))_{\text{KG}}$ with $\widetilde{\Phi}^-$ although we omit it because we will not use it in this paper.}
\begin{align}
     &\cO_{1+i\lambda}^-(z)=(\Phi^-_{1+i\lambda}(x;z),\phi^\tout(x))_{\text{KG}}, \quad 
    \cO_{1+i\lambda}^+(z)=(\Phi^+_{1+i\lambda}(x;z),\phi^\tin(x))_{\text{KG}},
    \\
    &\widetilde{\cO}_{1+i\lambda}^+(z)=(\widetilde{\Phi}^+_{1+i\lambda}(x;z),\phi^\tin(x))_{\text{KG}}
\end{align}
with massive conformal primary wave functions $\Phi^{\mp}_{1+i\lambda}(x;z)$ and $\widetilde{\Phi}^{+}_{1+i\lambda}(x;z)$ defined by
with 
\begin{align}
   \Phi^\mp_{1+i\lambda}(x;z)&=\mp \int [d\hp]\frac{ 2^{\mp i\lambda}m^{1\pm i\lambda} \sqrt{\lambda}G_{1+i\lambda}(\hp; z)}{(2\pi)^\frac{5}{2}}e^{\mp i m \hp\cdot x},
   \\
    % \Phi^-_{1+i\lambda}(x;z)&=- \int [d\hp]\frac{ 2^{-i\lambda}m^{1+i\lambda} \sqrt{\lambda}G_{1+i\lambda}(\hp; z)}{(2\pi)^\frac{5}{2}}e^{-i m \hp\cdot x},
    % \\
    % \Phi^+_{1+i\lambda}(x;z)&=\int [d\hp]\frac{2^{i\lambda} m^{1-i\lambda}\sqrt{\lambda}G_{1+i\lambda}(\hp; z)}{(2\pi)^\frac{5}{2}}e^{+i m \hp\cdot x},
    \widetilde{\Phi}^+_{1+i\lambda}(x;z)&=\int [d\hp]\frac{2^{i\lambda} m^{1-i\lambda}\sqrt{\lambda}G_{1-i\lambda}(\hp; z)}{(2\pi)^\frac{5}{2}}e^{+i m \hp\cdot x}.
\end{align}
The integrals can be performed, using the formula,\footnote{We need a regularization to 
 make the integral converge.}
\begin{align}
\int [d\hp] G_{1+i\lambda}(\hp; q)e^{\pm i m\hp \cdot x}
=\frac{4\pi e^{\mp \frac{i\pi}{2}}e^{\pm \frac{\pi \lambda}{2}}(\sqrt{x^2})^{i\lambda}}{m(-q\cdot x\mp i \epsilon)^{1+i\lambda}}K_{i\lambda}(m\sqrt{x^2}),
\end{align}
as
\begin{align}
\Phi^\mp_{1+i\lambda}(x;z)&=-\frac{i\, 2^{1\mp i\lambda}e^{\mp\frac{\pi}{2}\lambda}\sqrt{\lambda}m^{\pm i\lambda}(\sqrt{x^2})^{i\lambda}}{(2\pi)^\frac{3}{2}(-q(z)\cdot x)^{1\pm i\lambda}}K_{i\lambda}(m\sqrt{x^2}),
    \\
    \widetilde{\Phi}^+_{1+i\lambda}(x;z)&=-\frac{i\,2^{1+i\lambda}e^{-\frac{\pi}{2}\lambda}\sqrt{\lambda}(m\sqrt{x^2})^{-i\lambda}}{(2\pi)^\frac{3}{2}(-q(z)\cdot x)^{1-i\lambda}}K_{-i\lambda}(m\sqrt{x^2}).
\end{align}

%%%%%%%%%%
%%%%%%%%%%
\subsection{Massless limit}
Let us consider the massless limit of the massive conformal wave functions. 
Near the massless limit, the Bessel function is expanded as 
\begin{align}
    K_{i\lambda}(m\sqrt{x^2})=2^{-1+i\lambda}\Gamma(i\lambda)(m\sqrt{x^2})^{-i\lambda}+2^{-1-i\lambda}\Gamma(-i\lambda)(m\sqrt{x^2})^{i\lambda}+\mathcal{O}(m^2).
\end{align}
We thus have, near $m=0$, 
\begin{align}
    \Phi^-_{1+i\lambda}(x;z)&= -\frac{i\,e^{-\frac{\pi\lambda}{2}}\sqrt{\lambda}}{(2\pi)^\frac{3}{2}(-q(z)\cdot x)^{1+i\lambda}}
     \left[\Gamma(i\lambda)+ 2^{-2i\lambda}\Gamma(-i\lambda)(m\sqrt{x^2})^{2i\lambda}+\mathcal{O}(m^2)
    \right].
\end{align}
The Riemann–Lebesgue lemma ($\lim_{x\to \infty}e^{\pm i \lambda x}=0$ as a distribution) implies we can neglect the second oscillating term as $\lim_{m\to 0}m^{\pm 2i\lambda}=0$.
Therefore, the massless limit of the massive wave function $\Phi^-_{1+i\lambda}$ is
\begin{align}
   \lim_{m\to0} \Phi^-_{1+i\lambda}(x;z)&= -\frac{i\,e^{-\frac{\pi\lambda}{2}}\sqrt{\lambda}\Gamma(i\lambda)}{(2\pi)^\frac{3}{2}(-q(z)\cdot x)^{1+i\lambda}}=\varphi^-_{1+i\lambda}(x;z).
\end{align}
This exactly\footnote{To be honest, the phases of the massless conformal wave functions are arbitrary and we fixed them so that they agree with the massless limit of the massive wave functions.} agrees with massless wave functions $\varphi^-_{1+i\lambda}$ given by \eqref{vphi-}.
Similarly, we reproduce massless conformal wave functions \eqref{vphi+}, \eqref{tvphi+} as follows: 
\begin{align}
     \lim_{m\to0} \Phi^+_{1+i\lambda}(x;z)&= -\frac{i\,e^{\frac{\pi}{2}\lambda}\sqrt{\lambda}\Gamma(-i\lambda)(x^2)^{i\lambda}}{(2\pi)^\frac{3}{2}(-q(z)\cdot x)^{1-i\lambda}}=\varphi^+_{1+i\lambda}(x;z).
     \\
     \lim_{m\to0} \tilde{\Phi}^+_{1+i\lambda}(x;z)&= -\frac{i\,e^{-\frac{\pi\lambda}{2}}\sqrt{\lambda}\Gamma(-i\lambda)}{(2\pi)^\frac{3}{2}(-q(z)\cdot x)^{1-i\lambda}}=\widetilde{\varphi}^+_{1+i\lambda}(x;z).
\end{align}
Therefore, the celestial dictionary for massive scalars \eqref{massive-dict-amp} supports our rectified dictionary for massless one \eqref{massless-dict-amp}.

%%%%%%%%%%%%%
\subsection{Comments on crossing symmetry}
Crossing symmetry relates amplitudes involving an incoming positive energy particle to ones involving an outgoing negative energy particle.
% Crossing symmetry states the following two amplitudes are the same.\red{Can change?}
% \begin{itemize}
% \item one incoming positive energy particle, one outgoing positive energy amplitude
% \item one outgoing negative energy particle, one outgoing positive energy amplitude
% \end{itemize}
% The former has already been defined, so we would like to define the latter.
However, we have not explicitly established a dictionary for negative energy particles.
It would be natural to treat all outgoing particles on an equal footing regardless of whether they have positive or negative energy, as below:
\begin{align}
   \ev{\mathcal{O}_1^{-,m}(z_1)\mathcal{O}_{2}^{\prime-,m}(z_2)} \propto \int[dp_1][dp_2] G_{1+i\lambda_1}(p_1;z_1)G_{1+i\lambda_2}(-p_2;z_2)A(p_1,-p_2; \text{nothing}). \label{CR1}
\end{align}
Here, $p_2$ is a future-directed vector, and thus $-p_2$ is a past-directed one representing negative energy. 
We note that the superscript ``$-$'' on operators is used here to denote outgoing, irrespective of whether it carries positive or negative energy, and ``$m$'' represents that the operators correspond to massive particles in the Minkowski spacetime.
$\mathcal{O}^-$ is interpreted as operators for particles with outgoing positive energy, while $\mathcal{O}^{\prime-}$ is interpreted as those for particles with outgoing negative energy.

\begin{figure}[t]
  \centering
  \includegraphics[width=15cm]{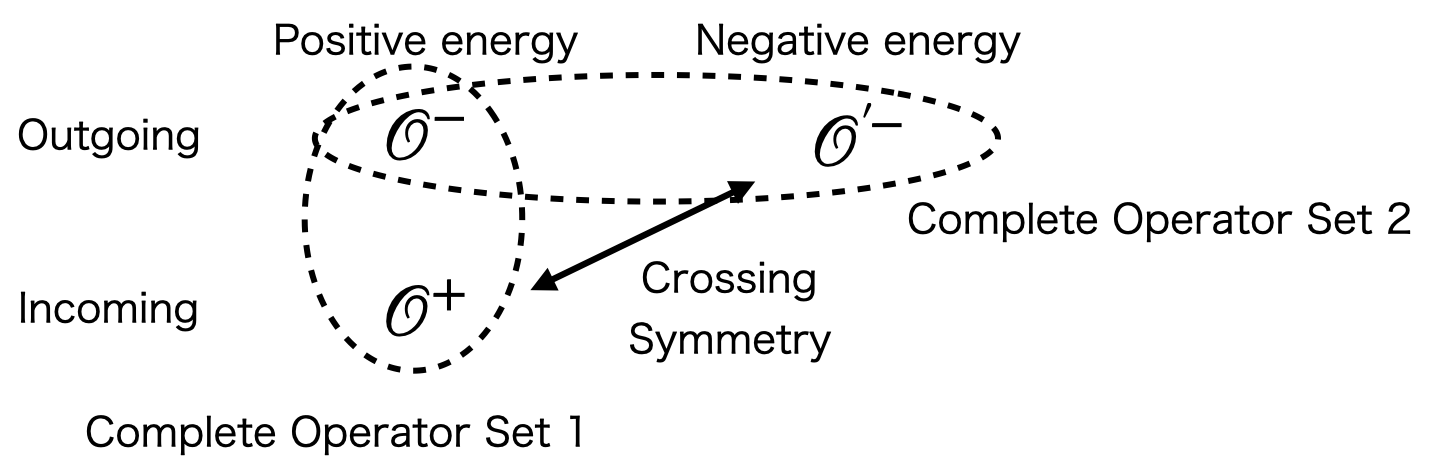}
  \caption{The relationship within the operator set in CCFT. Crossing symmetry establishes the equivalence between outgoing positive energy operators and incoming negative energy operators.}
  \label{OPSET}
\end{figure}

For comparison, we can write the amplitude again from the viewpoint of interpretation which includes one incoming positive energy particle and one outgoing positive energy particle,
\begin{align}
   \ev{\mathcal{O}_{1}^{-,m}(z_1)\mathcal{O}_{2}^{+,m}(z_2)}& \propto \int[dp_1][dp_2] G_{1+i\lambda_1}(p_1;z_1)G_{1+i\lambda_2}(p_2;z_2)A(p_1;p_2),\label{CR2} 
\end{align}
${\mathcal{O}}^{+}$ is interpreted as an operator for particles with incoming positive energy.
From crossing symmetry, the following equation holds:
\begin{align}
A(p_1,-p_2; \text{nothing}) = A(p_1;p_2).
\end{align}
From this relation, comparing \eqref{CR1} with \eqref{CR2}, we can establish the relation between $\mathcal{O}^{\prime -}$ and $\mathcal{O}^{+}$ as follows (see Fig.\ref{OPSET}):
\begin{align}
    \mathcal{O}_{2}^{\prime -,m}\propto
    \mathcal{O}_{2}^{+,m}.
\end{align} 
%%%%%%%%%
%%%%%%%%%
The above discussion holds even when considering $\mathcal{O}_1^-$ as more general operators in CCFT. In other words, regarding $\mathcal{O}_2$, the above relation holds at the operator level. 
%We have considered the massive fields as an example in the above. 

In our rectified dictionary, we determine that \eqref{CR1} takes the form of the standard two-point function in CFT in both massive and massless cases. 
In the massive case, \eqref{CR1} and \eqref{CR2} take an ordinary form of two-point functions in two-dimensional CFT. 
This is the reason why we do not have to modify the dictionary for massive fields up to normalization factors. 
Since we can smoothly take the massless limit as we have seen, a similar discussion should hold even for massless fields.\footnote{A situation is more subtle for massless scalars. A naive extension of the massless dictionary to the negative energy is
\begin{align}
  \ev{\mathcal{O}_1^-(z_1)\mathcal{O}_2^{\text{naive}~-}(z_2)} \propto \int d\omega_1 d\omega_2\omega^{i\lambda_1}(-\omega_2)^{i\lambda_2}A(p_1,-p_2; nothing).
\end{align}
However, this naive extension does not work well, and we need a more careful extension so that it is consistent with the massless limit.}

\section{Operator product expansion}\label{Sec3}
 The previous proposed celestial dictionary leads to a peculiar property in that the two-point functions take the delta function form while higher-point functions follow the power-law. In this section, we focus on OPE, making ensure the consistency of our definition in the dictionary.
Three-point functions in two-dimensional CFT take the form
\begin{align}
\ev{\mathcal{O}_1(z_1) \mathcal{O}_2(z_2) \mathcal{O}_3(z_3)} = \frac{f_{123}}{|z_{12}|^{\Delta_1+\Delta_2-\Delta_3}|z_{23}|^{\Delta_2+\Delta_3-\Delta_1}|z_{31}|^{\Delta_3+\Delta_1-\Delta_2}},
\label{OPE1}
\end{align}
where $f_{123}$ is a structure constant.
In two-dimensional CFT, the OPE is schematically expressed as (see e.g. \cite{Simmons-Duffin:2016gjk}):
\begin{align}
\mathcal{O}_1(z_1) \mathcal{O}_2(z_2) = \sum_{k} C_{12k}(z_{12}, \partial_1,\bar{\partial}_1)\mathcal{O}_k(z_1),
\end{align}
where $k$ runs over the primary operators in CCFT.\footnote{More precisely, the sum over $k$ should be an integral because the CCFT has a continuous spectrum.}
Utilizing this equation, the three-point function can be reformulated as follows:
\begin{align}
\ev{\mathcal{O}_1(z_1) \mathcal{O}_2(z_2) \mathcal{O}_3(z_3)} = \sum_{k} C_{12k}(z_{12}, \partial_1,\bar{\partial}_1)\ev{\mathcal{O}_k(z_1) \mathcal{O}_3(z_3)}.
\label{OPE2}
\end{align}
In the case $\ev{\mathcal{O}_k(z_1) \mathcal{O}_3(z_3)}\propto \frac{1}{|z_{13}|^{2\Delta_{3}}}$, we can write three-point function as follows:
\begin{align}
\ev{\mathcal{O}_1(z_1) \mathcal{O}_2(z_2) \mathcal{O}_3(z_3)} \propto \sum_{k}C_{12k}(z_{12}, \partial_1,\bar{\partial}_1)\frac{1}{|z_{13}|^{2\Delta_{3}}}. 
\label{OPE3}
\end{align}
On the other hand, in the case $\ev{\mathcal{O}_k(z_1) \mathcal{O}_3(z_3)}\propto \delta^{(2)}({z_1-z_3})$, the three-point function becomes
\begin{align}
 \ev{\mathcal{O}_1(z_1) \mathcal{O}_2(z_2) \mathcal{O}_3(z_3)} \propto \sum_k C_{12k}(z_{12}, \partial_1,\bar{\partial}_1)\delta^{(2)}({z_1-z_3}).
 \label{deltaOPE}
\end{align}
In \eqref{OPE1}, there is no distributional behavior. Thus we should have \eqref{OPE3} rather than \eqref{deltaOPE}. 
In the next section, we compute a concrete example for three-point functions.

\subsection{Two-massless and one-massive  three-point function}
Here, we consider a tree level scattering amplitude for two incoming massless and one outgoing massive scalar fields with the coupling constant $g$ (considered in \cite{Lam:2017ofc, Raclariu:2021zjz}), and reevaluate it by our dictionary.
The scattering amplitude is given by
\begin{align}
\mathcal{A}(1,2,3)=-i(2\pi)^4 g \delta^{(4)}(\omega_1 \hat{q}_1+\omega_2 \hat{q}_2- m \hat{p}_3).
\end{align}
In our dictionary, the corresponding celestial three-point function becomes
\begin{align}
    &\langle 
    \widetilde{\mathcal{O}}^+_{\Delta_1=1+i\lambda_1}(z_1) \widetilde{\mathcal{O}}^+_{\Delta_2=1+i\lambda_2}(z_2)\mathcal{O}^{-,m}_{\Delta_3=1+i\lambda_3} (z_3)
    \rangle\nonumber\\
    =&\prod_{i=1,2}\int d\omega_i \frac{i\omega_i^{-i\lambda}}{\sqrt{(2\pi)^3\lambda_i }} \int [d\hp_3] c(m,\lambda_3)G_{1+i\lambda_3}(\hp_3; z_3)i\mathcal{A}(1,2,3)\nonumber\\
    =&-g (2\pi)^{-\frac{3}{2}}\sqrt{\frac{\lambda_3}{\lambda_1\lambda_2}} \frac{m^{-\Delta_1-\Delta_2+\Delta_3}}{2^{-\Delta_1-\Delta_2+\Delta_3+3}} \frac{B\left(\frac{\Delta_{1}^\ast-\Delta_2^\ast+\Delta_3}{2}, \frac{\Delta_{2}^\ast-\Delta_1^\ast+\Delta_3}{2}\right)}{\left|z_{12}\right|^{\Delta_1^\ast+\Delta_2^\ast-\Delta_3}\left|z_{13}\right|^{\Delta_1^\ast+\Delta_3-\Delta_2^\ast}\left|z_{23}\right|^{\Delta_2^\ast+\Delta_3-\Delta_1^\ast}}.
\end{align}
Taking the inverse shadow transformation for the incoming particles,\footnote{
We use the following formula \cite{Simmons-Duffin:2012juh}:
\begin{align}
\int d^2z_0\frac{1}{|z_{10}|^{2a}|z_{20}|^{2b}|z_{30}|^{2c}}=\frac{2\pi\Gamma(1-a)\Gamma(1-b)\Gamma(1-c)}{\Gamma(a)\Gamma(b)\Gamma(c)}\frac{1}{|z_{12}|^{2(1-c)}|z_{23}|^{2(1-a)}|z_{31}|^{2(1-b)}}
\end{align}
where $z_{ij}\equiv z_i-z_j$.
It leads to 
\begin{align}
    &\frac{\Delta^{\ast}_1-1}{2\pi}\frac{\Delta^{\ast}_2-1}{2\pi}\int d^2z_1d^2z_2\frac{1}{|z_{11^{\prime}}|^{2\Delta_1}}\frac{1}{|z_{22^{\prime}}|^{2\Delta_2}} \frac{1}{\left|z_{12}\right|^{\Delta_1^\ast+\Delta_2^\ast-\Delta_3}\left|z_{13}\right|^{\Delta_1^\ast+\Delta_3-\Delta_2^\ast}\left|z_{23}\right|^{\Delta_2^\ast+\Delta_3-\Delta_1^\ast}}\nonumber\\
    %=&\frac{\Gamma(2-\Delta_1)\Gamma(2-\Delta_2)}{\Gamma(\Delta_1)\Gamma(\Delta_2)}\frac{\Gamma(\frac{\Delta_1+\Delta_2+\Delta_3}{2}-1)}{\Gamma(2-\frac{\Delta_1+\Delta_2+\Delta_3}{2})}
    %\frac{\Gamma(\frac{\Delta_1+\Delta_2-\Delta_3}{2})}{\Gamma(1-\frac{\Delta_1+\Delta_2-\Delta_3}{2})} \frac{1}{\left|z_{1^{\prime}2^{\prime}}\right|^{\Delta_1+\Delta_2-\Delta_3}\left|z_{1^{\prime}3}\right|^{\Delta_1+\Delta_3-\Delta_2}\left|z_{2^{\prime}3}\right|^{\Delta_2+\Delta_3-\Delta_1}}\nonumber\\
    =&\frac{\Delta_1^{\ast}+\Delta_2^{\ast}+\Delta_3-2}{\Delta_1+\Delta_2+\Delta_3-2}\frac{B(\Delta_1^{\ast},\Delta_2^{\ast})}{B(\Delta_1,\Delta_2)}
    \frac{B(\frac{\Delta_1+\Delta_2+\Delta_3}{2},\frac{\Delta_1+\Delta_2-\Delta_3}{2})}{B(\frac{\Delta_1^{\ast}+\Delta_2^{\ast}+\Delta_3}{2},\frac{\Delta_1^{\ast}+\Delta_2^{\ast}-\Delta_3}{2})}\frac{1}{\left|z_{1^{\prime}2^{\prime}}\right|^{\Delta_1+\Delta_2-\Delta_3}\left|z_{1^{\prime}3}\right|^{\Delta_1+\Delta_3-\Delta_2}\left|z_{2^{\prime}3}\right|^{\Delta_2+\Delta_3-\Delta_1}}.
\end{align}
}
we obtain
\begin{align}
    &\langle{\mathcal{O}}^+_{\Delta_1}(z_1) {\mathcal{O}}^+_{\Delta_2}(z_2)\mathcal{O}^{-,m}_{\Delta_3} (z_3)
    \rangle
    =\frac{C(\Delta_1,\Delta_2,\Delta_3)}{\left|z_{12}\right|^{\Delta_1+\Delta_2-\Delta_3}\left|z_{13}\right|^{\Delta_1+\Delta_3-\Delta_2}\left|z_{23}\right|^{\Delta_2+\Delta_3-\Delta_1}},
    \label{3pt-explicit}
\end{align}
where
\begin{align}
    C(\Delta_1,\Delta_2,\Delta_3)
    :=&-g (2\pi)^{-\frac{3}{2}}\sqrt{\frac{\lambda_3}{\lambda_1\lambda_2}} \frac{m^{-\Delta_1-\Delta_2+\Delta_3}}{2^{-\Delta_1-\Delta_2+\Delta_3+3}}\frac{{\Delta_1^{\ast}+\Delta_2^{\ast}+\Delta_3}-2}{{\Delta_1+\Delta_2+\Delta_3}-2}\nonumber\\
    &\times\frac{B\left(\frac{\Delta_{1}-\Delta_2+\Delta_3}{2}, \frac{\Delta_{2}-\Delta_1+\Delta_3}{2}\right)
    B(\frac{\Delta_1+\Delta_2+\Delta_3}{2},\frac{\Delta_1+\Delta_2-\Delta_3}{2})
    B(\Delta_1^{\ast},\Delta_2^{\ast})}{B(\frac{\Delta_1^{\ast}+\Delta_2^{\ast}+\Delta_3}{2},\frac{\Delta_1^{\ast}+\Delta_2^{\ast}-\Delta_3}{2})B(\Delta_1,\Delta_2)}.
\end{align}
Eq.~\eqref{3pt-explicit} explicitly shows that the shadow transformation results in the form of the three-point function determined by the conformal dimensions.
The behavior of this three-point function for  $|z_{23}|\to0$ is
\begin{align}
    \langle{\mathcal{O}}^+_{\Delta_1}(z_1) {\mathcal{O}}^+_{\Delta_2}(z_2)\mathcal{O}^{-,m}_{\Delta_3} (z_3)
    \rangle
    \to \frac{C(\Delta_1,\Delta_2,\Delta_3)}{|z_{23}|^{\Delta_2+\Delta_3-\Delta_1}}\frac{1}{|z_{12}|^{2\Delta_1}}.
    \label{3pt-ope-limit}
\end{align}
On the other hand, if we take OPE for $\mathcal{O}^+_{\Delta_2}\mathcal{O}^{-,m}_{\Delta_3}$, we have
\begin{align}
    \langle{\mathcal{O}}^+_{\Delta_1}(z_1) {\mathcal{O}}^+_{\Delta_2}(z_2)\mathcal{O}^{-,m}_{\Delta_3} (z_3)\rangle
    =\sum_kC_{23k}(z_{23},\partial_{z_2})\langle \mathcal{O}^{+}_{\Delta_1} (z_1){\mathcal{O}}^k_{\Delta_k}(z_2)\rangle.
\end{align}
We should note that the sum over $k$ runs over all operators for both massless and massive particles.
However, from the discussion on the two-point functions, only massless operators contribute to this expansion because $\mathcal{O}^{+}_{\Delta_1}$ is the operator for the massless field.
If the two-point function takes the delta function form, the three-point function has to be proportional to $\delta^{(2)}(z_1-z_2)$ in the OPE limit $|z_{23}|\to0$ which is inconsistent with \eqref{3pt-ope-limit}.
Thus, the previous celestial dictionary in the literature may have inconsistency. 
In our rectified dictionary, the two-point function follows the standard power-law and it is consistent with the behavior of the three-point function \eqref{3pt-ope-limit} as in the usual CFTs.

\subsection{Four-point function}
As stated by \cite{Arkani-Hamed:2020gyp, Chang:2022jut}, 
the four-point celestial correlators for massless particles result in delta functions if we use the previous dictionary.
It is caused by the delta function form of the two-point functions from the following OPE:
\begin{align}
    \ev{\mathcal{O}_1(z_1)\mathcal{O}_2(z_2)\mathcal{O}_3(z_3)\mathcal{O}_4(z_4)}=\sum_{k,l}C_{12k}(z_{12},\partial_{z_1})C_{34l}(z_{34},\partial_{z_3})\ev{\mathcal{O}_k(z_1)\mathcal{O}_l(z_3)}.
\end{align}
It is obvious that, if the two-point function is proportional to the delta function, the four-point function exhibits distributional behavior like the delta function.
In our definition, we do not have such behavior and obtain
\begin{align}
    \ev{\mathcal{O}_1(z_1)\mathcal{O}_2(z_2)\mathcal{O}_3(z_3)\mathcal{O}_4(z_4)}=\sum_{k}C_{12k}(z_{12},\partial_{z_1})C_{34k}(z_{34},\partial_{z_3})\frac{1}{|z_1-z_3|^{2\Delta_k}}.
\end{align}

\section{Summary}\label{Sec4}
In this paper, we have rectified the holographic dictionary in the celestial holography so that the celestial two-point functions follow the standard power-law instead of the delta function.
The dictionary for massive scalars is smoothly connected to the one for massless scalars in the massless limit. 
We have also discussed that the dictionary is natural, as it is suggested by massless limit and crossing symmetry.
In our definition, the hermitian conjugate with respect to the four-dimensional Klein-Gordon inner product is the same as the shadow transformation in CCFT.
Furthermore, the behavior of the OPE according to this definition is consistent with the results of the two-point function.
Our redefinition has led to a substantial reevaluation of properties in CCFT, specifically in OPEs and conformal block expansion  \cite{Lam:2017ofc, Atanasov:2021cje, Fan:2021isc}. 

\section*{Acknowledgements}
We are grateful to Ana-Maria Raclariu,  Kotaro Tamaoka, Seiji Terashima, and Takashi Tsuda for useful discussions.
HF is supported by JSPS Grant-in-Aid for Scientific Research KAKENHI Grant No. JP22J14381.
The work of NO and TW was supported by JST, the establishment of university fellowships towards the creation of science technology innovation, Grant Number
JPMJFS2123.
The work of NO is partially supported by Grant-inAid for Transformative Research Areas (A) “Extreme Universe” No. 21H05187.
SS acknowledges support from JSPS KAKENHI Grant Numbers JP21K13927 and JP22H05115.

\bibliographystyle{utphys}
\bibliography{ref-celes}

\providecommand{\href}[2]{#2}\begingroup\raggedright\begin{thebibliography}{10}

\bibitem{tHooft:1993dmi}
G.~'t~Hooft, ``{Dimensional reduction in quantum gravity},'' {\em Conf. Proc. C} {\bfseries 930308} (1993) 284--296, \href{http://arxiv.org/abs/gr-qc/9310026}{{\ttfamily arXiv:gr-qc/9310026}}.

\bibitem{Susskind:1994vu}
L.~Susskind, ``{The World as a hologram},'' \href{http://dx.doi.org/10.1063/1.531249}{{\em J. Math. Phys.} {\bfseries 36} (1995) 6377--6396}, \href{http://arxiv.org/abs/hep-th/9409089}{{\ttfamily arXiv:hep-th/9409089}}.

\bibitem{Maldacena:1997re}
J.~M. Maldacena, ``{The Large N limit of superconformal field theories and supergravity},'' \href{http://dx.doi.org/10.4310/ATMP.1998.v2.n2.a1}{{\em Adv. Theor. Math. Phys.} {\bfseries 2} (1998) 231--252}, \href{http://arxiv.org/abs/hep-th/9711200}{{\ttfamily arXiv:hep-th/9711200}}.

\bibitem{Gubser:1998bc}
S.~S. Gubser, I.~R. Klebanov, and A.~M. Polyakov, ``{Gauge theory correlators from noncritical string theory},'' \href{http://dx.doi.org/10.1016/S0370-2693(98)00377-3}{{\em Phys. Lett. B} {\bfseries 428} (1998) 105--114}, \href{http://arxiv.org/abs/hep-th/9802109}{{\ttfamily arXiv:hep-th/9802109}}.

\bibitem{Witten:1998qj}
E.~Witten, ``{Anti-de Sitter space and holography},'' \href{http://dx.doi.org/10.4310/ATMP.1998.v2.n2.a2}{{\em Adv. Theor. Math. Phys.} {\bfseries 2} (1998) 253--291}, \href{http://arxiv.org/abs/hep-th/9802150}{{\ttfamily arXiv:hep-th/9802150}}.

\bibitem{Strominger:2001pn}
A.~Strominger, ``{The dS / CFT correspondence},'' \href{http://dx.doi.org/10.1088/1126-6708/2001/10/034}{{\em JHEP} {\bfseries 10} (2001) 034}, \href{http://arxiv.org/abs/hep-th/0106113}{{\ttfamily arXiv:hep-th/0106113}}.

\bibitem{Maldacena:2002vr}
J.~M. Maldacena, ``{Non-Gaussian features of primordial fluctuations in single field inflationary models},'' \href{http://dx.doi.org/10.1088/1126-6708/2003/05/013}{{\em JHEP} {\bfseries 05} (2003) 013}, \href{http://arxiv.org/abs/astro-ph/0210603}{{\ttfamily arXiv:astro-ph/0210603}}.

\bibitem{Pasterski:2016qvg}
S.~Pasterski, S.-H. Shao, and A.~Strominger, ``{Flat Space Amplitudes and Conformal Symmetry of the Celestial Sphere},'' \href{http://dx.doi.org/10.1103/PhysRevD.96.065026}{{\em Phys. Rev. D} {\bfseries 96} no.~6, (2017) 065026}, \href{http://arxiv.org/abs/1701.00049}{{\ttfamily arXiv:1701.00049 [hep-th]}}.

\bibitem{Cardona:2017keg}
C.~Cardona and Y.-t. Huang, ``{S-matrix singularities and CFT correlation functions},'' \href{http://dx.doi.org/10.1007/JHEP08(2017)133}{{\em JHEP} {\bfseries 08} (2017) 133}, \href{http://arxiv.org/abs/1702.03283}{{\ttfamily arXiv:1702.03283 [hep-th]}}.

\bibitem{Strominger:2017zoo}
A.~Strominger, {\em {Lectures on the Infrared Structure of Gravity and Gauge Theory}}.
\newblock 3, 2017.
\newblock \href{http://arxiv.org/abs/1703.05448}{{\ttfamily arXiv:1703.05448 [hep-th]}}.

\bibitem{Pasterski:2017kqt}
S.~Pasterski and S.-H. Shao, ``{Conformal basis for flat space amplitudes},'' \href{http://dx.doi.org/10.1103/PhysRevD.96.065022}{{\em Phys. Rev. D} {\bfseries 96} no.~6, (2017) 065022}, \href{http://arxiv.org/abs/1705.01027}{{\ttfamily arXiv:1705.01027 [hep-th]}}.

\bibitem{Pasterski:2017ylz}
S.~Pasterski, S.-H. Shao, and A.~Strominger, ``{Gluon Amplitudes as 2d Conformal Correlators},'' \href{http://dx.doi.org/10.1103/PhysRevD.96.085006}{{\em Phys. Rev. D} {\bfseries 96} no.~8, (2017) 085006}, \href{http://arxiv.org/abs/1706.03917}{{\ttfamily arXiv:1706.03917 [hep-th]}}.

\bibitem{Donnay:2020guq}
L.~Donnay, S.~Pasterski, and A.~Puhm, ``{Asymptotic Symmetries and Celestial CFT},'' \href{http://dx.doi.org/10.1007/JHEP09(2020)176}{{\em JHEP} {\bfseries 09} (2020) 176}, \href{http://arxiv.org/abs/2005.08990}{{\ttfamily arXiv:2005.08990 [hep-th]}}.

\bibitem{Ogawa:2022fhy}
N.~Ogawa, T.~Takayanagi, T.~Tsuda, and T.~Waki, ``{Wedge holography in flat space and celestial holography},'' \href{http://dx.doi.org/10.1103/PhysRevD.107.026001}{{\em Phys. Rev. D} {\bfseries 107} no.~2, (2023) 026001}, \href{http://arxiv.org/abs/2207.06735}{{\ttfamily arXiv:2207.06735 [hep-th]}}.

\bibitem{Kapec:2014opa}
D.~Kapec, V.~Lysov, S.~Pasterski, and A.~Strominger, ``{Semiclassical Virasoro symmetry of the quantum gravity $ \mathcal{S}$-matrix},'' \href{http://dx.doi.org/10.1007/JHEP08(2014)058}{{\em JHEP} {\bfseries 08} (2014) 058}, \href{http://arxiv.org/abs/1406.3312}{{\ttfamily arXiv:1406.3312 [hep-th]}}.

\bibitem{Kapec:2016jld}
D.~Kapec, P.~Mitra, A.-M. Raclariu, and A.~Strominger, ``{2D Stress Tensor for 4D Gravity},'' \href{http://dx.doi.org/10.1103/PhysRevLett.119.121601}{{\em Phys. Rev. Lett.} {\bfseries 119} no.~12, (2017) 121601}, \href{http://arxiv.org/abs/1609.00282}{{\ttfamily arXiv:1609.00282 [hep-th]}}.

\bibitem{Pasterski:2021rjz}
S.~Pasterski, ``{Lectures on celestial amplitudes},'' \href{http://dx.doi.org/10.1140/epjc/s10052-021-09846-7}{{\em Eur. Phys. J. C} {\bfseries 81} no.~12, (2021) 1062}, \href{http://arxiv.org/abs/2108.04801}{{\ttfamily arXiv:2108.04801 [hep-th]}}.

\bibitem{Raclariu:2021zjz}
A.-M. Raclariu, ``{Lectures on Celestial Holography},'' \href{http://arxiv.org/abs/2107.02075}{{\ttfamily arXiv:2107.02075 [hep-th]}}.

\bibitem{Crawley:2021ivb}
E.~Crawley, N.~Miller, S.~A. Narayanan, and A.~Strominger, ``{State-operator correspondence in celestial conformal field theory},'' \href{http://dx.doi.org/10.1007/JHEP09(2021)132}{{\em JHEP} {\bfseries 09} (2021) 132}, \href{http://arxiv.org/abs/2105.00331}{{\ttfamily arXiv:2105.00331 [hep-th]}}.

\bibitem{Chang:2022jut}
C.-M. Chang, W.~Cui, W.-J. Ma, H.~Shu, and H.~Zou, ``{Shadow celestial amplitudes},'' \href{http://dx.doi.org/10.1007/JHEP02(2023)017}{{\em JHEP} {\bfseries 02} (2023) 017}, \href{http://arxiv.org/abs/2210.04725}{{\ttfamily arXiv:2210.04725 [hep-th]}}.

\bibitem{Chang:2022seh}
C.-M. Chang and W.-J. Ma, ``{Missing corner in the sky: massless three-point celestial amplitudes},'' \href{http://dx.doi.org/10.1007/JHEP04(2023)051}{{\em JHEP} {\bfseries 04} (2023) 051}, \href{http://arxiv.org/abs/2212.07025}{{\ttfamily arXiv:2212.07025 [hep-th]}}.

\bibitem{He:2014cra}
T.~He, P.~Mitra, A.~P. Porfyriadis, and A.~Strominger, ``{New Symmetries of Massless QED},'' \href{http://dx.doi.org/10.1007/JHEP10(2014)112}{{\em JHEP} {\bfseries 10} (2014) 112}, \href{http://arxiv.org/abs/1407.3789}{{\ttfamily arXiv:1407.3789 [hep-th]}}.

\bibitem{Polchinski:1998rq}
J.~Polchinski, \href{http://dx.doi.org/10.1017/CBO9780511816079}{{\em {String theory. Vol. 1: An introduction to the bosonic string}}}.
\newblock Cambridge Monographs on Mathematical Physics. Cambridge University Press, 12, 2007.

\bibitem{Simmons-Duffin:2012juh}
D.~Simmons-Duffin, ``{Projectors, Shadows, and Conformal Blocks},'' \href{http://dx.doi.org/10.1007/JHEP04(2014)146}{{\em JHEP} {\bfseries 04} (2014) 146}, \href{http://arxiv.org/abs/1204.3894}{{\ttfamily arXiv:1204.3894 [hep-th]}}.

\bibitem{Simmons-Duffin:2016gjk}
D.~Simmons-Duffin, \href{http://dx.doi.org/10.1142/9789813149441_0001}{``{The Conformal Bootstrap},''} in {\em {Theoretical Advanced Study Institute in Elementary Particle Physics}: {New Frontiers in Fields and Strings}}, pp.~1--74.
\newblock 2017.
\newblock \href{http://arxiv.org/abs/1602.07982}{{\ttfamily arXiv:1602.07982 [hep-th]}}.

\bibitem{Lam:2017ofc}
H.~T. Lam and S.-H. Shao, ``{Conformal Basis, Optical Theorem, and the Bulk Point Singularity},'' \href{http://dx.doi.org/10.1103/PhysRevD.98.025020}{{\em Phys. Rev. D} {\bfseries 98} no.~2, (2018) 025020}, \href{http://arxiv.org/abs/1711.06138}{{\ttfamily arXiv:1711.06138 [hep-th]}}.

\bibitem{Arkani-Hamed:2020gyp}
N.~Arkani-Hamed, M.~Pate, A.-M. Raclariu, and A.~Strominger, ``{Celestial amplitudes from UV to IR},'' \href{http://dx.doi.org/10.1007/JHEP08(2021)062}{{\em JHEP} {\bfseries 08} (2021) 062}, \href{http://arxiv.org/abs/2012.04208}{{\ttfamily arXiv:2012.04208 [hep-th]}}.

\bibitem{Atanasov:2021cje}
A.~Atanasov, W.~Melton, A.-M. Raclariu, and A.~Strominger, ``{Conformal block expansion in celestial CFT},'' \href{http://dx.doi.org/10.1103/PhysRevD.104.126033}{{\em Phys. Rev. D} {\bfseries 104} no.~12, (2021) 126033}, \href{http://arxiv.org/abs/2104.13432}{{\ttfamily arXiv:2104.13432 [hep-th]}}.

\bibitem{Fan:2021isc}
W.~Fan, A.~Fotopoulos, S.~Stieberger, T.~R. Taylor, and B.~Zhu, ``{Conformal blocks from celestial gluon amplitudes},'' \href{http://dx.doi.org/10.1007/JHEP05(2021)170}{{\em JHEP} {\bfseries 05} (2021) 170}, \href{http://arxiv.org/abs/2103.04420}{{\ttfamily arXiv:2103.04420 [hep-th]}}.

\end{thebibliography}\endgroup

\end{document}